\documentclass[twocolumn,showpacs,preprintnumbers,amsmath,amssymb]{revtex4-2}
\usepackage{graphicx}
\usepackage{dcolumn}
\usepackage{bm}
\usepackage{subfigure}
\usepackage[colorlinks,linkcolor=blue,anchorcolor=blue,citecolor=blue]{hyperref}
\expandafter\let\csname equation*\endcsname\relax\expandafter\let\csname endequation*\endcsname\relax\usepackage{amsmath}
\usepackage{diagbox}
\usepackage{mfirstuc}
\usepackage{enumerate}
\usepackage{makecell}

\begin{document}
\preprint{APS/123-QED}

\title{Quantum Secret Reconstruction}

\author{Ruihai Ma$^{1}$}
\author{Fei Gao$^{2}$}%
\altaffiliation{Corresponding author. Email address: gaof@bupt.edu.cn}
\author{Song Lin$^{1}$}%
\altaffiliation{Corresponding author. Email address: lins95@gmail.com}
\affiliation{$^1$ College of Computer and Cyber Security, Fujian Normal University, Fuzhou, 350117, China\\$^2$State Key Laboratory of Networking and Switching Technology, Beijing University of Posts and Telecommunications, Beijing, 100876, China}%

\date{\today}

\begin{abstract}
In addition to secret splitting, secret reconstruction is another important component of secret sharing. In this paper, the first quantum secret reconstruction protocol based on cluster states is proposed. Before the protocol, a classical secret is divided into multiple shares, which are distributed among shareholders via secret splitting. In the protocol, the dealer utilizes her secret to encrypt a private quantum state, and sends the encrypted state to a combiner chosen by her from the shareholders. With the help of other shareholders, the combiner utilizes the properties of cluster states to recover the privacy quantum state. It is shown that the proposed protocol is secure against several common attacks, including external and internal attacks. Compared with classical secret reconstruction protocols, this protocol not only achieves theoretical security of all shares, but also is more efficient due to reducing the distribution cost and computation cost. To demonstrate the feasibility of the protocol, a corresponding simulation quantum experiment is conducted on the IBM Q platform. Furthermore, in conjunction with quantum fingerprinting, it can be directly applied to achieve the task of multiple secrets sharing, because the classical shares can be reused in the proposed protocol.  
\end{abstract}


\maketitle

\section{\label{sec1}Introduction}
Secret sharing (SS) was proposed by Blakley\cite {Blakley1979} and Shamir\cite {shamir} respectively in 1979, which has become one of the most important techniques in cryptographic primitives. So far, SS has been associated with a range of applications, such as online auctions, electronic voting, collaborative businesses and so on. However, with the emergence of quantum algorithms, classical secret sharing, the security of which is based on computational complexity, is not secure any more. Aiming for enhancing the security of SS, Hillery et al$.$ proposed the first quantum secret sharing (QSS) protocol \cite{qss1} in 1999, which exploited the correlation of Greenberger-Horne-Zeilinger states \cite{ghz}. Since then, an increasing number of quantum protocols have been proposed based on different techniques, such as quantum entanglement swapping \cite{qes1,qes2,qes3}, quantum Fourier transform \cite{Yang2013,Mashhadi2019,Qin2018,Song2017,Sutradhar2020} and so on.

In a simple secret sharing, a secret is first split into two shares and they are distributed to two shareholders. After that, these two shareholders can utilize their shares to recover the secret only when they collaborate. Evidently, a complete secret sharing consists of two processes: secret splitting and secret reconstruction. However, the existing quantum secret sharing protocols focus on secret splitting, and ignore the reconstruction process. In fact, the protection of both splitting and reconstruction processes has been studied in classical SS. During the classic secret reconstruction, it is necessary to transmit large amounts of classical information in the secure communication channel. Only a portion of the shares is available for transmission on public channels \cite{Wang2008,Beimel1998-compare}. Moreover, the identity authentication with public key cryptosystems is also necessary in the classical reconstruction. 

To solve the above issues, we propose the quantum secret reconstruction protocol  with cluster states  that satisfies the information theoretical security in this paper. To the best of our knowledge, the proposed protocol is the first quantum secret reconstruction protocol. In the proposed protocol, the reconstructed secret is only accessible to the combiner, but not to outsiders and the other participants. The protocol emphasizes the protection of each shareholder's distributed classical share. Neither dishonest shareholders nor external attackers are able to steal shares in the protocol. In the protocol, the cluster state acts as a carrier of information, pooling all the shares together, and eventually recovering the privacy quantum state. The protocol not only affords security advantages over the classical secret reconstruction protocols, but also the efficiency advantage of utilizing distributed shares and computing the corresponding data. As well, since all shares are perfectly hidden, which implies that the shares are all reusable. In combination with quantum fingerprinting technique \cite{Buhrman2001}, the protocol can further implement a feasible multiple secret sharing scheme.

The rest of the paper is organized as follows. In Sec.II, we briefly review the cluster state and its property. Then, the protocol process is described in Sec.III. The correction, security and efficiency of the proposed protocol are analyzed in Sec.IV. A simulation experiment and a quantum multiple secrets sharing scheme are provided in Sec.V and VI respectively. At last, a conclusion is given in Sec.VII.

\section{Preliminary}
\begin{figure}[!t]
	\centering
	\includegraphics[width=1.75in]{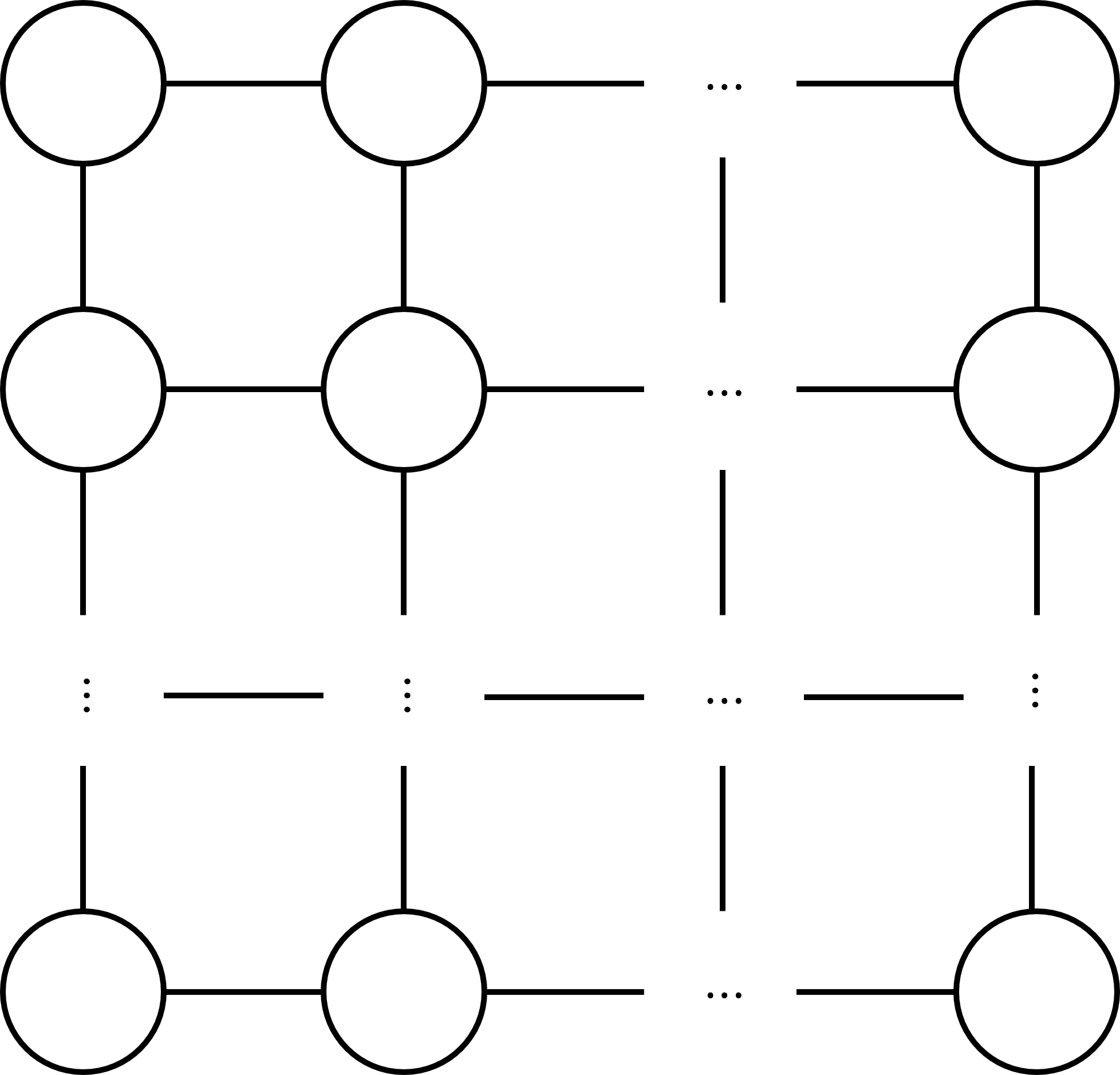}
	\caption{Graphical representation of the cluster state.}
	\label{fig1}
\end{figure}

As one of the most representative entangled states, cluster states\cite{cluster1,Nielsen2006,cluster3,cluster4,cluster5,cluster6,cluster7} are widely used in quantum cryptography, quantum computing and other fields. It can be represented in the form of graphs, denoted by $G(V,E)$, as shown in FIG. \ref{fig1}. Here, $V$ is the set of all nodes in graph $G$, and $E$ is the set of all edges. In graph G, each node represents a qubit in the state $|+\rangle=H|0\rangle=\frac{1}{\sqrt{2}}(|0\rangle+|1\rangle)$, and each edge represents a controlled-phase operation $CZ$, where $H$ and $CZ$ operation are shown in the following equations.

\begin{equation}\label{HCZ}
	\begin{aligned}
		H=\frac{1}{\sqrt{2}}\begin{bmatrix}
			1 & 1  \\
			1 & -1 
		\end{bmatrix}, \\
		CZ=\begin{bmatrix}
			1 & 0 & 0 & 0 \\
			0 & 1 & 0 & 0 \\
			0 & 0 & 1 & 0 \\
			0 & 0& 0 & -1 
		\end{bmatrix}. 
	\end{aligned}
\end{equation}
The cluster state $|C\rangle$ has some favorable properties for secret reconstruction and satisfies the following characteristic equation.
\begin{equation}
	K |C \rangle = |C\rangle,
	\label{eigen_equation}
\end{equation}

\noindent where $K = X_{a} \mathop{\otimes}\limits_{b\in N(a)} Z_{b}\mathop{\otimes}\limits_{c \in (V-a-N(a))} I_c$ is the stabilizer operator for
\begin{equation}
	\begin{aligned}
		X&=\begin{bmatrix}
			0 & 1 \\
			1 & 0
		\end{bmatrix},
		Z=\begin{bmatrix}
			1 & 0 \\
			0 & -1
		\end{bmatrix},
		I=\begin{bmatrix}
			1 & 0 \\
			0 & 1
		\end{bmatrix}.\\
	\end{aligned}
	\label{IXZ}
\end{equation}
The subscript $a$ denotes an arbitrary node in the cluster state, $N(a)$ denotes the neighboring nodes set of node $a$, and $c \in (V-a-N(a))$ denotes the other nodes in the cluster state. 

In addition, $R_X(\theta)$ and $R_Z(\theta)$ are the rotation operations corresponding to $X$ and $Z$, respectively, as shown in the following equations, where $\theta$ is an arbitrary rotation angle.
\begin{equation}
	\begin{aligned}
		R_X(\theta)&=\begin{bmatrix}
			cos\frac{\theta}{2} & -isin\frac{\theta}{2} \\
			-isin\frac{\theta}{2} & cos\frac{\theta}{2}
		\end{bmatrix},
		R_Z(\theta)=\begin{bmatrix}
			e^{-i\frac{\theta}{2}} & 0 \\
			0 & e^{+i\frac{\theta}{2}}
		\end{bmatrix}.
	\end{aligned}
	\label{rxz}
\end{equation}
\begin{figure}[!t]
	\centering
	\includegraphics[width=2.2 in]{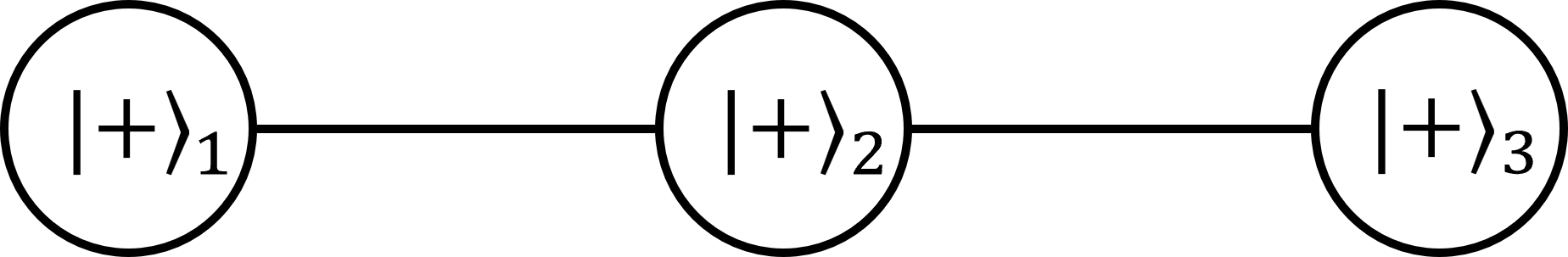}
	\caption{The example represents the three-particle cluster state $|C\rangle_{1,2,3}$ with the initial state represented in the node.}
	\label{fig2}
\end{figure}
For clarity, we will take the preparation of a three-particle cluster state $|C\rangle_{1,2,3}$ as an example, where subscripts denote the order of the particles. We can start by preparing three particles in the state $|+\rangle_1$, $|+\rangle_2$ and $|+\rangle_3$. And then, we perform CZ operations between $|+\rangle_1$ and $|+\rangle_2$, $|+\rangle_2$ and $|+\rangle_3$, respectively. The $|C\rangle_{1,2,3}$ represented in FIG. \ref{fig2} is the state
\begin{equation}
	\begin{aligned}
		|C\rangle_{1,2,3}&=(I_1\otimes CZ_{2,3})\;(CZ_{1,2}\otimes I_3) (|+\rangle_1\otimes|+\rangle_2\otimes|+\rangle_3)\\
		&=\frac{1}{\sqrt{2}}(I\otimes CZ_{2,3})(|0\rangle_1|+\rangle_2+|1\rangle_1|-\rangle_2)\otimes|+\rangle_3\\
		&=\frac{1}{\sqrt{2}}(|+\rangle_1 |0\rangle_2 |+\rangle_3 + |-\rangle_1 |1\rangle_2 |-\rangle_3).
	\end{aligned}	
	\label{CZCZ}
\end{equation}
\noindent and is the unique common eigenstate of
\begin{equation}	\label{ce}
	\begin{aligned}
		X_1Z_2I_3|C\rangle_{1,2,3}=|C\rangle_{1,2,3},\\
		Z_1X_2Z_3|C\rangle_{1,2,3}=|C\rangle_{1,2,3},\\
		I_1Z_2X_3|C\rangle_{1,2,3}=|C\rangle_{1,2,3}.
	\end{aligned}
\end{equation}
Based on these properties, we construct the first quantum secret reconstruction protocol in the following paper.

\section{Protocol}
Before the protocol, the dealer, Alice, has the flexibility to choose from different protocols, such as QSS, classical secret sharing, quantum key distribution protocols and others, to share a secret with $n$ shareholders. These shareholders include individuals such as Bob$_1$, Bob$_2$, $\cdots$, Bob$_{n-1}$ and Bob$_n$. That is, the dealer Alice holds a secret $k_A$, and each shareholder Bob$_i$ holds a share $k_i$, where $i=1,2,\cdots,n$. These values satisfy the follow equation.
\begin{equation}
	\centering
	k_A+\sum_{i=1}^{n}k_i\;\rm{mod}\;q=0,
	\label{key-eq}
\end{equation}
\noindent where $\rm{q}$ is a prime number greater than 2. Based on this, Alice desires to share a privacy quantum state $|\psi\rangle_A = \alpha|0\rangle + \beta|1\rangle$, where $|\alpha|^2 + |\beta|^2 = 1$. And she also hopes that the privacy of the secret and shares throughout the reconstruction process is ensured. Alice randomly selects a shareholder, referred to as the combiner, to reconstruct the privacy quantum state $|\psi\rangle$. Without loss of generality, let Bob$_n$ be the combiner who reconstructs $|\psi\rangle$, and for convenience, we refer to Bob$_n$ as Charlie and his share as $k_C = k_n$.

\subsection{Protocol Process}
\begin{figure}[!t]
	\centering
	\includegraphics[width=3.5 in]{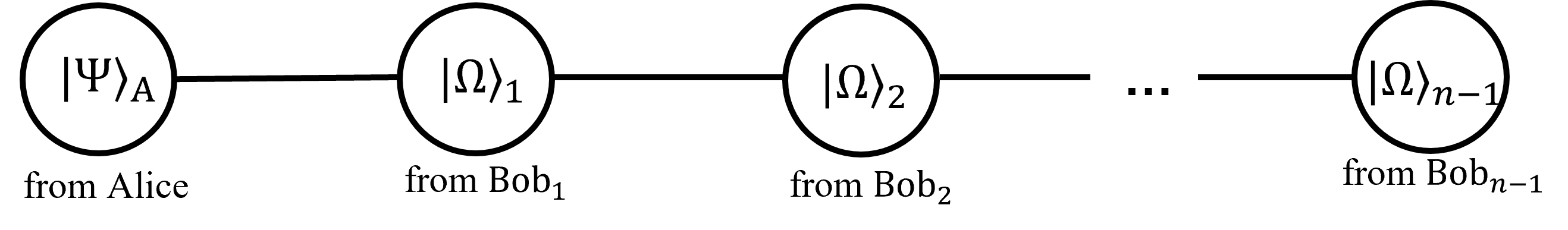}
	\caption{The $n$-particle one-dimensional cluster state in the proposed protocol.}
	\label{fig3}
\end{figure}

\noindent \textbf{(S$_1$)} Alice randomly chooses a number $s\in \{1,2,\cdots,$ $ {q-1}\}$ and informs all the shareholders. Then, Alice, Bob$_1$, Bob$_2$, $\cdots$, Bob$_{n-1}$ and Charlie compute the rotation angles $\phi_j$ with their secret and shares $k_j$, where $j\in\{A,1,2,\cdots,n-1,C\}$, according to
\begin{equation}
	\phi_j=s\cdot \frac{k_j}{\rm{q}}\cdot2\pi.
	\label{encode-eq}
\end{equation} 
Then, Alice encrypts $|\psi\rangle_A$ to $|\Psi\rangle_A=R_X(\phi_A)|\psi\rangle_A$ with the angle $\phi_A$. Obviously, all the rotation angles satisfies the following equation,
\begin{equation}
	\phi_A+\sum_{i=1}^{n-1}\phi_i+\phi_C = 2\pi r.
	\label{phi_sum}
\end{equation} 
\noindent where $r$ denotes an integer. 
\\\\
\noindent\textbf{(S$_2$)} Alice applies the decoy method to transmit $|\Psi\rangle_A$ to Charlie through the quantum secure channel. In other words, she sends particles in the BB84 states \cite{bb84} along with $|\Psi\rangle_A$ to detect any potential eavesdropping.
\\\\
\noindent \textbf{(S$_3$)} Each of Bob$_1$, Bob$_2$, $\cdots$, Bob$_{n-1}$ selects a random angle $\omega_i$ and prepares $|\Omega\rangle{_i}=R_Z(\omega_i)|+\rangle=|+_{\omega_i}\rangle$, where $i=1,2,\cdots n-1$. The angle $\omega_i$ is only known to Bob$_i$ and not to any other person. Every Bob then sends his particle to Charlie through the quantum secure channel either.
\\\\
\noindent\textbf{(S$_4$)} After receiving $|\Psi\rangle_A$ and $|\Omega\rangle_{i}$, Charlie constructs a $n$-particle one-dimensional cluster state. He performs CZ operations on each pair of adjacent particles in sequence. Alice's particle $|\Psi\rangle_A$ is placed in the first position, refer to as particle A, and Bob$_i$'s particle $|\Omega\rangle{_i}$ is placed in the $(i+1)$ position, refer to as particle $i$, as shown in FIG. \ref{fig3}.
\\\\
\noindent\textbf{(S$_5$)} Charlie measures the particle A in the $\{|+\rangle, |-\rangle\}$ basis. If the outcome is $|+\rangle$, the measurement result is $m_1=0$. If the outcome is $|-\rangle$, the measurement result is $m_1=1$. He transmits the measurement result $m_1$ to Bob$_1$. Meanwhile, Charlie applies a compensating operation to the particle 1 according to $m_1$. If $m_1=0$, Charlie applies $I$ compensating operation to the particle 1. If $m_1=1$, Charlie applies $X$ compensating operation to the particle 1.
\\\\
\noindent\textbf{(S$_6$)} Bob$_1$ calculates a new rotation angle $\theta_1=(-1)^{m_1+1}\omega_1+\phi_1$ based on Charlie's announced measurement result $m_1$, and informs Charlie. Then, Charlie applies the operation $R_X(\theta_1)H$ to the particle 1.
\\\\
\noindent \textbf{(S$_{2i+3}$,$\,i$$=$$2,3,\cdots,{n-1}$)} Charlie measures the particle ${i-1}$ in the $\{|+\rangle,|-\rangle\}$ basis and obtains the measurement result $m_i$. He transmits $m_i$ to Bob$_i$ and applies a compensating operation to the particle $i$. If $m_i=0$, Charlie applies $I$ compensating operation. If $m_i=1$, Charlie applies $X$ compensating operation.
\\\\
\noindent \textbf{(S$_{2i+4}$,$\,i$$=$$2,3,\cdots,{n-1}$)} Bob$_i$ calculates the rotation angle $\theta_i=(-1)^{m_i+1}\omega_i+\phi_i$ based on the measurement result $m_i$, and informs Charlie. Charlie then applies the $R_X(\theta_i)H$ operation to the particle $i$.
\\\\
\noindent \textbf{(S$_{2n+3}$)} Finally, Charlie obtains a quantum state on the last qubit in the form of the following equation,
\begin{equation}
	|\Psi'\rangle=R_X(\phi_A+\sum_{i=1}^{n-1}\phi_i)|\psi\rangle.
	\label{last-quantum-state}
\end{equation}
\noindent According to Eq.(\ref{phi_sum}), by applying the $R_X(\phi_{C})$ operation to the particle $|\Psi'\rangle$, Charlie will recover the privacy quantum state $|\psi\rangle$, disregarding the global phase..

\subsection{A simple example}
To be clear, we will provide an example in this section. Suppose there is a dealer named Alice and three shareholders named Bob$_1$, Bob$_2$ and Charlie for secret reconstruction. Alice desires to share a privacy quantum state $|\psi\rangle_A=\frac{1}{2}|0\rangle+\frac{\sqrt{3}}{2}|1\rangle$ and Charlie is the combiner. Without loss of generality, suppose that the secret of the dealer and the shares of three shareholders are $k_A=2$, $k_1=1$, $k_2=2$ and $k_C=1$. Moreover, $s=1$ and $\rm{q}=3$.

\begin{figure}[!t]
	\centering
	\includegraphics[width=2.2 in]{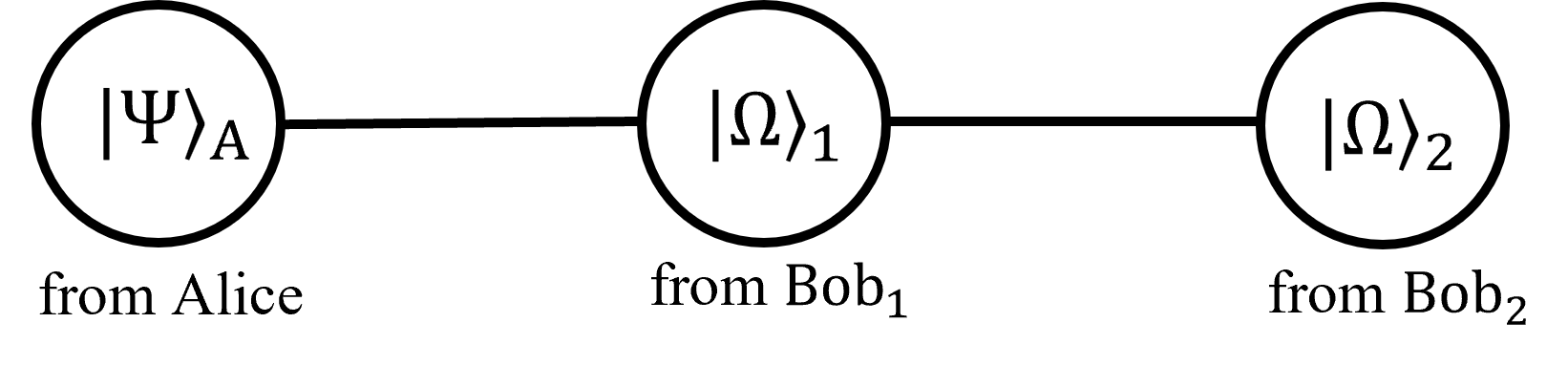}
	\caption{The cluster state $|C\rangle$ is constructed by entangling particles $|\Psi\rangle_A$, $|\Omega\rangle_1$ and $|\Omega\rangle_2$.}
	\label{fig4}
\end{figure}

Firstly, all participants calculate the rotation angles $\phi$ according to Eq.(\ref{encode-eq}). It can be obtained that $\phi_A=2\pi/3$, $\phi_1=\pi/3$, $\phi_2=2\pi/3$ and $\phi_C=\pi/3$. Next, Alice encrypts the privacy quantum state $|\psi\rangle_A$ with $\phi_A=2\pi/3$ to obtain $|\Psi\rangle_A=R_X(2\pi/3)|\psi\rangle_A=\frac{1-3i}{4}|0\rangle+\frac{\sqrt{3}-\sqrt{3}i}{4}|1\rangle$. And she transmits it to Charlie through the quantum secure channel. Bob$_1$ chooses a random angle $\omega_1=\pi/6$, prepares $|\Omega\rangle_1=|+_{\pi/6}\rangle$ and transmits it to Charlie through the quantum secure channel. Similarly, Bob$_2$ chooses $\omega_2=\pi$, prepares $|\Omega\rangle_2=|+_{\pi}\rangle$ and then transmits it to Charlie through the quantum secure channel. After receiving particles $|\Psi\rangle_A$, $|\Omega\rangle_1$ and $|\Omega\rangle_2$, Charlie constructs the cluster state $|C\rangle$, shown in FIG. \ref{fig4}. Each particle of $|C\rangle$ is connected by $CZ$, and the quantum state $|C\rangle$ is shown in the following equation,

\begin{equation}
	\begin{aligned}
		|C\rangle_{A,1,2}=&(I\otimes CZ_{1,2})(CZ_{A,1}\otimes I)(|\Psi\rangle_A \otimes|\Omega\rangle_1\otimes|\Omega\rangle_2)\\
		=&\frac{1}{\sqrt{2}}(\frac{1-3i}{4}(e^{-i\frac{\pi}{12}}|00+_{\pi}\rangle+e^{+i\frac{\pi}{12}}|01-_{\pi}\rangle)_{A,1,2}+ \\ &\frac{\sqrt{3}-\sqrt{3}i}{4}(e^{-i\frac{\pi}{12}}|10+_{\pi}\rangle-e^{+i\frac{\pi}{12}}|11-_{\pi}\rangle)_{A,1,2}).
	\end{aligned}	
	\label{cs-e}
\end{equation}

Secondly, Charlie begins to measure and perform corresponding operations on these particles. Without loss of generality, we suppose that the two measurement results of particle A and particle 1 are 0 and 1, respectively. Charlie measures the particle A in the $\{|+\rangle,|-\rangle\}$ basis and obtains a measurement result $m_1=0$. He then transmits the result $m_1=0$ to Bob$_1$ and performs the compensating operation $I$ on the particle 1. Bob$_1$ calculates the rotation angle $\theta_1=-\omega_1+\phi_1=\pi/6$ and tells Charlie. Charlie performs $R_X(\pi/6)H$ operation on the particle 1, and the state of the three particles becomes $|+\rangle_A\otimes e^{-i\frac{\pi}{2}}(\frac{\sqrt{3}}{2}|0+_{\pi}\rangle+\frac{1}{2}|1-_{\pi}\rangle)_{1,2}$.

Thirdly, Charlie measures the particle 1 in the $\{|+\rangle,|-\rangle\}$ basis and obtains a measurement result $m_2=1$. He tells Bob $_2$ the measurement $m_2=1$ and performs the compensating operation. Since $m_2=1$, Charlie performs $X$ compensating operation on the particle 2. Bob$_2$ then calculates $\theta_2=2\pi/3+\pi=5\pi/3$ based on $m_2=1$ and tells Charlie. Then Charlie performs $R_X(5\pi/3)H$ operation on the particle 2. At this point, Charlie obtains the quantum state $|\Psi'\rangle=
\frac{-\sqrt{3}-\sqrt{3}i}{4}|0\rangle+\frac{-3-i}{4}|1\rangle)_2$ on the particle 2.

Finally, Charlie performs $R_X(\phi_C)=R_X(\pi/3)$ operation on $|\Psi'\rangle$. And he will recover the privacy quantum state $|\psi\rangle_A=\frac{1}{2}|0\rangle+\frac{\sqrt{3}}{2}|1\rangle$, disregarding the global phase. The concrete process of the above example is shown in Table \ref{te}.

\begin{table*}[!t]
	\caption{Concrete process of the example\label{te}}
	\centering
		\begin{ruledtabular}
		\begin{tabular}{cc}
			Concrete process of the example & Quantum states of the three particles \\
			\hline
			\makecell[c]{Alice, Bob$_1$ and Bob$_2$ \\transmit particles $|\Psi\rangle_A, |+_{\pi/6}\rangle_1$ and $ |+_{\pi}\rangle_2$} & \makecell[c]{$(\frac{1-3i}{4}|0\rangle+\frac{\sqrt{3}-\sqrt{3}i}{4}|1\rangle)_A\otimes$$|+_{\pi/6}\rangle_1\otimes|+_{\pi}\rangle_2$}\\
			\hline
			\makecell[c]{Charlie entangles these three \\particles with the controlled-phase \\operations $CZ_{A,1}$ and $CZ_{1,2}$} &\makecell[c]{$\frac{1}{\sqrt{2}}(\frac{1-3i}{4}(e^{-i\frac{\pi}{12}}|00+_{\pi}\rangle+e^{+i\frac{\pi}{12}}|01-_{\pi}\rangle)_{A,1,2}$ \\$+ $ $\frac{\sqrt{3}-\sqrt{3}i}{4}(e^{-i\frac{\pi}{12}}|10+_{\pi}\rangle-e^{+i\frac{\pi}{12}}|11-_{\pi}\rangle)_{A,1,2})$} \\
			\hline
			\makecell[c]{Charlie measures the first particle \\in the $\{|+\rangle,|-\rangle\}$ basis,\\ performs the compensating operation \\and tells Bob$_1$ the result $m_1=0$} & \makecell[c]{$|+\rangle_A\otimes\frac{1}{\sqrt{2}}(\frac{1-3i}{4}(e^{-i\frac{\pi}{12}}|0+_{\pi}\rangle+e^{+i\frac{\pi}{12}}|1-_{\pi}\rangle)+$\\$\frac{\sqrt{3}-\sqrt{3}i}{4}(e^{-i\frac{\pi}{12}}|0+_{\pi}\rangle-e^{+i\frac{\pi}{12}}|1-_{\pi}\rangle))_{1,2}$}\\
			\hline
			\makecell[c]{Bob$_1$ announces the angle $\theta_1=\pi/6$ \\and Charlie performs the \\operation $R_X(\pi/6)_1H_1$ on the second particle} & \makecell[c]{$|+\rangle_A\otimes e^{-i\frac{\pi}{2}}(\frac{\sqrt{3}}{2}|0+_{\pi}\rangle+\frac{1}{2}|1-_{\pi}\rangle)_{1,2}$} \\
			\hline
			\makecell[c]{Charlie measures the second particle \\in the $\{|+\rangle,|-\rangle\}$ basis,\\ performs the compensating operation \\and tells Bob$_2$ the result $m_2=1$}
			& \makecell[c]{$|+\rangle_A\otimes|-\rangle_1\otimes 
			e^{-i\frac{\pi}{2}}	(\frac{\sqrt{3}}{2}|+_{\pi}\rangle-\frac{1}{2}|-_{\pi}\rangle)_2$}\\
			\hline
			\makecell[c]{Bob$_2$ announces the angle $\theta_2=5\pi/3$ \\and Charlie performs the \\operation $R_X(5\pi/3)_2H_2$ on the third particle} 
			& \makecell[c]{$|+\rangle_A\otimes|-\rangle_1\otimes(\frac{-\sqrt{3}-\sqrt{3}i}{4}|0\rangle+\frac{-3-i}{4}|1\rangle)_2$}\\
			\hline \makecell{Charlie performs the $R_X(\pi/3)_2$ \\operation on the third particle}  & \makecell[c]{$|+\rangle_A\otimes|-\rangle_1\otimes
				e^{-i\pi}(\frac{1}{2}|0\rangle+\frac{\sqrt{3}}{2}|1\rangle)_2$}\\
		\end{tabular}%
	    \end{ruledtabular}
\end{table*}

\section{Performance analysis}
\subsection{Correctness}
Before proving the correctness of the proposed protocol, we first study the properties of two-particle cluster states. Take the two-particle cluster state $|C\rangle_{1,2}$ as an example. The $|C\rangle_{1,2}$ is created by entangling an arbitrary quantum state  $|\varphi\rangle_1=\alpha|0\rangle+\beta|1\rangle$ and a quantum state $|+_{\omega}\rangle_2$ with an arbitrary rotation angle $\omega$ using the $CZ$ operation, where $\alpha$ and $\beta$ are complex numbers that satisfy $|\alpha|^2+|\beta|^2=1$, and $|+_{\omega}\rangle_2=R_Z(\omega)|+\rangle$. In addition, $|-_{\omega}\rangle=R_Z(\omega)|-\rangle=\frac{1}{\sqrt{2}}(e^{-i\frac{\omega}{2}}|0\rangle-e^{+i\frac{\omega}{2}}|1\rangle)$. We can express $|C\rangle_{1,2}$ in the following equation,
\begin{equation}
	\begin{aligned}
		|C\rangle_{1,2}=&CZ_{1,2}|\varphi\rangle_1|+_\omega\rangle_2\\
		=&\alpha|0\rangle_1|+_\omega\rangle_2 + \beta|1\rangle_1|-_\omega\rangle_2\\
		=&\frac{1}{\sqrt{2}}|+_\theta\rangle_1\otimes(R_Z(\omega)HR_Z(-\theta)|\varphi\rangle)_2+\\
		&\frac{1}{\sqrt{2}}|-_\theta\rangle_1\otimes(R_Z(\omega)XHR_Z(-\theta)|\varphi\rangle)_2.
	\end{aligned}	
	\label{2pcs-1}
\end{equation}

\noindent According to Eq.(\ref{2pcs-1}), we can find that after measuring the first particle of $|C\rangle_{1,2}$ in the $\{|+_\theta\rangle,|-_\theta\rangle\}$ basis, we will obtain a quantum state $R_Z(\omega)X^{m}HR_Z(\theta)|\varphi\rangle$ on the second particle, where $\theta$ is an arbitrary measuring angle and $m$ represents the measurement result of the first particle.

In Ref.\cite{Nielsen2006}, Nielsen et al. proved that a multi-particle one-dimensional cluster state is equivalent to the combination of multiple two-particle cluster states, which delays the execution of $CZ$ operation until after the measurement. In other words, when we entangle and measure a multi-particle cluster state, we can first entangle two particles to form a two-particle cluster state. Then, we can measure the first particle of the two-particle cluster state and obtain an output state on the second particle. We use this output state as a new input state, entangle it with the next particle, and form a new two-particle cluster state. This process is repeated until the last particle is reached. The above steps are equivalent to entangling all particles at once and then measuring them one by one. Our proof will be based on this property.

We will now prove the correctness of the proposed protocol. According to the conclusion of Ref.\cite{Nielsen2006}, in steps \textbf{S$_{2i+3}$} to \textbf{S$_{2i+4}$} of our protocol, where $i$ runs in $\{1,2,3,\cdots,n-1\}$, Charlie only needs to entangle two particles in each round of the loop rather than entangle all particles at once. We will prove that one round of the loop is equivalent to performing $R_X(\phi_i)$ operation on the input state $|\varphi_i\rangle$ of each round. When $i=1$, the initial input state is $|\varphi_1\rangle=|\Psi\rangle_A$.

At the start, Charlie entangles the input state $|\varphi_i\rangle$ with $|\Omega\rangle_i$ using the $CZ$ operation. As an example, for $i=1$, Charlie entangles $|\Psi\rangle_A$ and $|\Omega\rangle_1$ with each other using the $CZ$ operation. Then, Charlie measures the first particle in the $\{|+\rangle,|-\rangle\}$ basis, and obtains the output state $R_Z(\omega_i)X^{m_i}H|\varphi_i\rangle$, where $m_i\in\{0,1\}$ is the result of this measurement. Charlie then performs $X^{m_i}$ compensating operation on the output state. 

Specifically, when $m_i=0$, Charlie performs the $I$ operation, and when $m_i=1$, Charlie performs the $X$ operation. After the compensating operation, the output state becomes $X^{m_i}R_Z(\omega_i)X^{m_i}H|\varphi_i\rangle=R_Z((-1)^{m_i}\omega_i)H|\varphi\rangle$. Bob$_i$ then announces the rotation angle $\theta_i=(-1)^{m_i+1}\omega_i+\phi_i$, and Charlie performs $R_X(\theta_i)H$ operation on the output state, which becomes $R_X((-1)^{m_i+1}\omega_i+\phi_i)HR_Z((-1)^{m_i}\omega_i)H|\varphi\rangle=R_X(\phi_i)|\varphi_i\rangle$. The output quantum state $R_X(\phi_i)|\varphi_i\rangle$ will be used as the new input state $|\varphi_{i+1}\rangle$ in the next round.

Therefore, when the above loop reaches the last round, it is equivalent to performing $R_X(\sum_{i=1}^{n-1}\phi_i)$ rotation on the initial input state $|\Psi\rangle_A$. Charlie can obtain the quantum state $|\Psi'\rangle=R_X(\sum_{i=1}^{n-1}\phi_i)|\Psi\rangle=R_X(\phi_A+\sum_{i=1}^{n-1}\phi_i)|\psi\rangle$ on the last particle. Finally, Charlie performs $R_X(\phi_C)$ operation on $|\Psi'\rangle$, and according to Eq.(\ref{phi_sum}), he can correctly reconstruct the privacy quantum state $|\psi\rangle$, disregarding the global phase. Therefore, the proposed protocol satisfies correctness.

\subsection{Security}
In this section, we will discuss the robustness of the proposed protocol to external and internal attacks. We will focus on the internal attack from the combiner, as this attack poses the greatest threat to the protocol. The analysis will prove that the proposed protocol is robust against various common attacks.

\subsubsection{External Attack}
Suppose there is an external eavesdropper, Eve, who has two eavesdropping targets. One target is Alice's privacy quantum state $|\psi\rangle_A$, and the other is the shares $k_i$ and $k_C$ of the shareholders Bob$_i$ and Charlie, where $i$$=$$1,2,\cdots,n-1$. First, consider the case where Eve eavesdrops on Alice's privacy quantum state $|\psi\rangle_A$. Since Alice chooses a quantum secure channel to transmit the particles $|\Psi\rangle_A$, along with many decoy particles such as $|+\rangle$, $|-\rangle$, $|0\rangle$, $|1\rangle$. Any attack by Eve will inevitably introduce errors and be detected. Suppose that Eve doesn't mind being detected and intercepts $|\Psi\rangle_A$. In this case, since $|\Psi\rangle_A=R_X(\phi_A)|\psi\rangle_A$, and Alice does not disclose any information about $\phi_A$, Eve cannot eavesdrop on $|\psi\rangle_A$.

Next, consider the case where Eve eavesdrops on the shares $k_i$ and $k_C$ belonging to Bob$_i$ and Charlie. Eve has the capability to intercept the particles $|\Omega\rangle_i$ in the quantum secure channel. However, as $|\Omega\rangle_i$ does not contain any valuable information, Eve's eavesdropping becomes futile and only leads to the introduction of errors, resulting in the termination of the protocol. Therefore, for a rational Eve, it is more advantageous to focus on eavesdropping on the public information. The public information that Eve can access comprises the measurement results $m_i$ announced by Charlie and the rotation angles $\theta_i$ announced by Bob$_i$, where $i=1,2,\cdots, n-1$. 

In consideration of the above public information, Eve is able to deduce the sign of $\omega_i$ from the measurement result $m_i$, yet she cannot acquire any helpful information as $\theta_i$ is the sum of $\phi_i$ and $(-1)^{m_i}\omega_i$. The angle $\omega_i$ is hidden in $|\Omega\rangle_i$ and only known to Bob$_i$. Therefore, Eve cannot obtain $\phi_i$ from $\theta_i$, and she cannot infer $k_i$. Furthermore, if Eve eavesdrops on Charlie's share $k_C$, Charlie does not disclose any information other than the measurement result in this protocol. Hence, Eve is unable to eavesdrop on Charlie's share $k_C$. Overall, external eavesdropper Eve cannot obtain any useful information and the proposed protocol is secure against the external attack.

\subsubsection{Internal Attack}
In this section, we will analyze two types of internal attacks: the attack from the combiner and the collusion attacks. Regarding collusion attacks, we will consider two extreme cases where only one participant is honest. The first case is where only the combiner is honest, referred to as collusion attack I. The second case is where the combiner is dishonest, and only one of the other shareholders is honest, referred to as collusion attack II. We will prove that our protocol is secure against these above attacks.

\paragraph{Attack from the Combiner}
Compared to the external attack, the internal attack from the combiner, Charlie, poses a greater threat to the proposed protocol. Since Charlie has access to all the particles transmitted by the dealer Alice and the shareholders Bob$_i$, such as $|\Psi\rangle_A$ and $|\Omega\rangle_i$, where $i=1,2,\cdots,n-1$. Moreover, since Charlie is an internal participant, his eavesdropping behavior cannot be detected by the quantum secure channel. Therefore, Charlie can fully utilize the particles he possesses to eavesdrop on Alice's privacy quantum state $|\psi\rangle_A$ and Bob$_i$'s share $k_i$. In this case, we use Charlie* to represent Charlie as an eavesdropper.

First, let's consider the case where Charlie* attempts to eavesdrop on Alice's privacy quantum state $|\Psi\rangle_A=R_X(\phi_A)|\psi\rangle_A$. Since Alice does not disclose any information, Charlie* is unaware of Alice's rotation angle $\phi_A$. As a result, Charlie* cannot perform the necessary $R_X(-\phi_A)$ operation on $|\Psi\rangle_A$ to obtain $|\psi\rangle$.
Second, let's consider the case where Charlie eavesdrops on Bob$_i$'s share $k_i$. After receiving the particles from Alice and Bob$_i$, Charlie* can randomly publish a set of fake measurement results such as $m'_i \in \{0,1\}$ instead of following the protocol as usual. Here, $i=1,2,\cdots, {n-1}$. In this way,  Charlie* manipulates Bob$_i$ to announce the rotation angle $\theta'_i=(-1)^{m'_i+1}\omega_i+\phi_i$. Next, Charlie* performs the $R_Z(\theta'_i)X^{m'_i}$ operation on $|\Omega\rangle_i=R_Z(\omega_i)|+\rangle$, so that he can obtain the quantum states only containing the $\phi_i$ angle, as shown in the following equation,
\begin{equation}
	|\Omega'\rangle_i=R_Z(\theta'_i)X^{m'_i}R_Z(\omega_i)|+\rangle=R_Z(\phi_i)|+\rangle.
	\label{theft-e}
\end{equation}

After obtaining the above particles, Charlie* aims to distinguish these particles in order to obtain different shares $k_i$ of Bob$_i$. Charlie* may attempt to distinguish $|\Omega'\rangle_i$ through measurements, but in this way, Charlie* cannot obtain any useful information. This is because, according to Eq.(\ref{encode-eq}), the angles $\phi_i$ between each participant are non-orthogonal, such as $\pi/3$ and $2\pi/3$. If Charlie* performs measurement on $|\Omega'\rangle_i$, he will obtain results of either 0 or 1 each time. Therefore, Charlie* cannot obtain any information about the quantum state and, consequently, cannot obtain the share $k_i$ of Bob$_i$. Furthermore, when Charlie* measures $|\Omega'\rangle_i$, he will lose the information of $\phi_i$, rendering him unable to reconstruct $|\psi\rangle_A$. For a rational Charlie*, this would clearly be disadvantageous. Additionally, each time Alice shares a privacy quantum state, she publishes a different $s$, which prevents Charlie* from collecting multiple identical $|\Omega'\rangle_i$ to distinguish $\phi_i$. In summary, the proposed protocol is secure against the attack from the combiner.

\paragraph{Collusion Attack I}

In this section, we will discuss the case where only the combiner is honest, while all Bobs are dishonest. In this case, the colluding attackers, referred to as Bob$^*$s, attempt to eavesdrop on Charlie's share $k_C$ and the privacy quantum state $|\psi\rangle$. Let's first consider the case where Bob$^*$s attempt to eavesdrop on Charlie's share $k_C$. In the protocol, Charlie only reveals the measurement results without disclosing any additional information. Therefore, based on this, Bob$^*$s cannot eavesdrop on $k_C$. If Bob$^*$s want to eavesdrop on the privacy quantum state $|\psi\rangle$, they can intercept the particles $|\Psi\rangle_A$ sent by Alice. Then, Bob$^*$s apply rotations using their own angles $\phi_i$ to $|\Psi\rangle_A$ and obtain the following quantum state, where $i=1,2,\cdots,n-1$.
\begin{equation}
	|\Psi''\rangle_A=R_X(\phi_A+\phi_1+\phi_2+\cdots+\phi_{n-1})|\psi\rangle_A.
	\label{bob-theft}
\end{equation}
\noindent However, since Bob$^*$s lack Charlie's angle $\phi_C$, they will not be able to obtain Alice's secret $|\psi\rangle$. Therefore the colluding eavesdroppers cannot obtain any useful information, and the protocol is secure when only the combiner is honest.

\paragraph{Collusion Attack II}
Another case of collusion attack is that only one shareholder other than the combiner is honest. Without loss of generality, let's suppose that only Bob$_1$ is honest. Bob$^*_2$, Bob$^*_3$, $\cdots$, Bob$^*_{n-1}$ are dishonest and collude with Charlie* to eavesdrop on the share $k_1$ of Bob$_1$ and privacy quantum state $|\psi\rangle$ of Alice. In this case, the protocol can be regarded as a three-party protocol among Alice, Bob$_1$, and Charlie*, which is equivalent to the attack from the combiner discussed earlier. We can conclude that regardless of whether the combiner, Charlie*, attempts to eavesdrop on Bob$_1$'s share $k_1$ or Alice's privacy quantum state $|\psi\rangle_A$, he cannot obtain any useful information. Therefore, the protocol is secure when only one shareholder is honest. Furthermore, when the number of honest shareholders is greater than one, the proposed protocol is obviously more secure.

\subsubsection{Summary of Security Analysis}
In the above security analysis, we analyzed several eavesdropping attacks based on the different targets, such as the privacy quantum state and the shares. These attacks include an external attack and a number of internal attacks. Among them, the most threatening attack is the one from the combiner. However, the combiner is unable to obtain any useful information. It has been proven that our protocol is secure against these attacks.

\subsection{Efficiency}
Since no previous work has been done on quantum secret reconstruction protocols, we compare the protocol with some classical secret reconstruction protocols, such as Ref.\cite{Beimel1998-compare} and Ref.\cite{Harn2020-compare}. We will compare them in two folds, one is the size of the shares distributed, referred as to the distribution cost, and the other is the computation time required of the protocol, referred as to the computation cost. 

Specifically, the distribution cost is the total bits of the shares divided by the number of the shared information, such as the number of privacy quantum states in this protocol. And the computation cost is the total time used for computation divided by the number of the shared information. In the following comparison, we use $n$ to represent the number of participants, $m$ to represent the number of privacy quantum states, $|q|$ to represent the size of each classical share, $T_a$ to represent the time used for a single addition, and $T_m$ to represent the time used for a single multiplication. In addition, for $(t,n)$ threshold protocols or multivariate protocols, since the number $t$ or the number of variables is often of the same order of magnitude as $n$, i.e. $O(t)=O(n)$, for the sake of uniformity, we use $O(n)$ substitutes $O(t)$.

We will now calculate the distribution cost and computation cost of the presented protocol. The advantage of our protocol is the ability to share multiple numbers of information using the same set of shares, referred to as the share reuse. Which signifies that the protocol is theoretically capable of sharing an infinite number, i.e. ${m\rightarrow \infty}$, of private quantum states by distributing shares only once. In our protocol, each of the $n$ shareholders owns a $|q|$ bits share, and the total number of bits of shares is $n|q|$. Therefore the distribution cost of the protocol can be calculated as $DC=n|q|/m\rightarrow O(1)$. 

The computation cost of our protocol can be divided into two parts, i.e$.$ the cost of angles $\phi$ and the cost of angles $\theta$. The calculation of the angles $\phi$, as given in Eq.(\ref{encode-eq}), require three multiplication operations at each computation. However, each time a private quantum state is shared, only the value of $s$ will be changed. Therefore, the computation cost of each angle $\phi$ only includes $(2+m)$ multiplication operations. Hence, the computation cost of the angles $\phi$ can be expressed as $CC_{\phi}=(2+m)T_m/m=O(1)T_m$. When Bob$_i$ calculates the angle $\theta_i=\pm\omega_i+\phi_i$, he needs to perform one addition operation, where $i=1,2,\cdots,n-1$. In which the cost of negating $\omega_i$ is extremely low and we do not consider it. The computation cost of all angles $\theta_1,\theta_2,\cdots,\theta_{n-1}$ requires $(n-1)m$ addition operations. Therefore the computation cost of angles $\theta$ is $CC_{\theta}=(n-1)mT_a/m=O(n)T_a$. In summary, the total computation cost of this protocol is $CC=CC_{\phi}+CC_{\theta}=O(n)T_a+O(1)T_m$.

In the unrestricted scheme of \cite{Beimel1998-compare}, the distribution cost is $DC_1=O(|q|n^2)$, and since this protocol uses Shamir's \cite{shamir} sharing scheme, the computation cost is $CC_1=O(n)T_a+O(n^3)T_m$. In the basic secure secret reconstruction (SSR) scheme of \cite{Harn2020-compare}, the distribution cost is $DC_2=O(|q|n)$, and the computation cost is $CC_2=O(n)T_a+O(n^3)T_m$. In the bivariate scheme of \cite{Harn2020-compare}, the distribution cost is $DC_3=O(|q|n^2)$, and the computation cost is $CC_3=O(n)T_a+O(n^4)T_m$. The comparison of these protocols is shown in Table \ref{table-compare}. By comparison, it is evident that the proposed protocol offers advantages in terms of both distribution cost and computation cost, based on the ability to reuse shares.

\begin{table*}[!t]
	\caption{The Comparison of Protocols \label{table-compare}}
	\centering
	\resizebox{1\textwidth}{!}{%
		\begin{ruledtabular}
		\begin{tabular}{cccc}
			Protocol &  Distribution Cost & Computation Cost & Share Reuse\\
			\hline
			the proposed protocol & $O(1)$ & $O(n)T_a+O(1)T_m$ & Yes\\
			
			unrestricted scheme of \cite{Beimel1998-compare} & $O(|q|n^2)$ & $O(n)T_a+O(n^3)T_m$ & No \\
			
			basic SSR scheme of \cite{Harn2020-compare} & $O(\,|q|n\,)$ & $O(n)T_a+O(n^3)T_m$ & No \\
			
			bivariate scheme of \cite{Harn2020-compare} & $O(|q|n^2)$ & $O(n)T_a+O(n^4)T_m$ & No  \\
			
		\end{tabular}
		\end{ruledtabular}
	}
\end{table*}

\section{Experiment}
\begin{figure*}[!t]
	\centering
	
	\includegraphics[width=7.2 in]{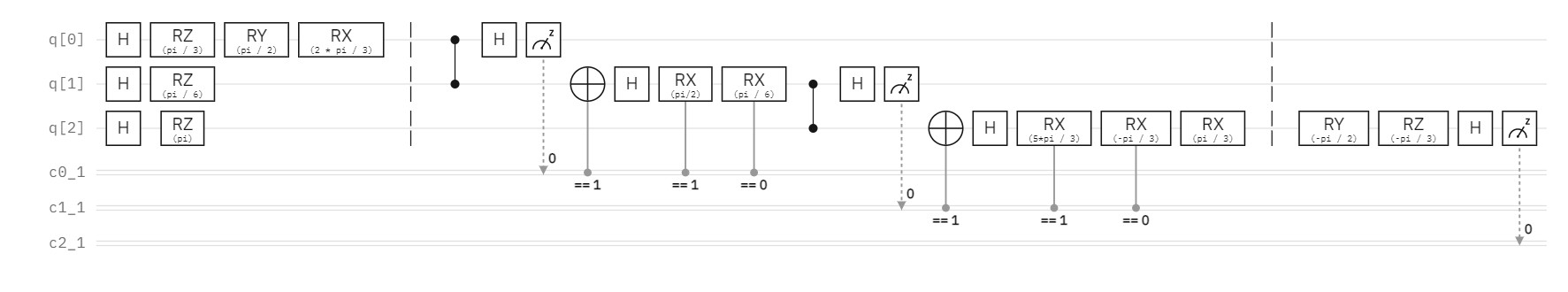}
	\caption{The circuit of the experiment.}
	\label{circuit}
\end{figure*}

To verify the feasibility of the protocol, we executed an experiment on the IBM Q platform with the job number chnfm6c2b9sdqn1jlojg, which everyone can check. For simplicity, the example in section III.B is utilized, the corresponding circuit is shown in FIG. \ref{circuit}. We will introduce the concrete implement in the following.

Firstly, we initialize the particles of every participants. To generate the quantum state $|\psi\rangle_A=\frac{1}{2}|0\rangle+\frac{\sqrt{3}}{2}|1\rangle$, we use the $H$, $R_Z(\pi/3)$ and $R_Y(\pi/2)$ gates on the $\rm{q}[0]$ circuit. Then we use the $H$ and $R_Z(\pi/6)$ gates on the $\rm{q}[1]$ circuit to generate $|\Omega\rangle_1=|+_{\pi/6}\rangle$. And we use the $H$ and $R_Z(\pi)$ gates on the $\rm{q}[2]$ circuit to generate $|\Omega\rangle_1=|+_{\pi}\rangle$. Next, we perform the $R_X(2\pi/3)$ operation on the $\rm{q}[0]$ circuit to simulate Alice encrypting $|\psi\rangle_A$ to $|\Psi\rangle_A=R_X(2\pi/3)|\psi\rangle_A$. 

Secondly, we begin to entangle and measure. To entangle the first and second circuit, we perform the $CZ$ operation between the $\rm{q[0]}$ and $\rm{q}[1]$ circuits. Since the IBM Q platform does not have the $\{|+\rangle,|-\rangle\}$ basis measurement, we use the $H$ gate and the $\{|0\rangle,|1\rangle\}$ basis measurement to achieve the equivalent measurement in the $\{|+\rangle,|-\rangle\}$ basis. The measurement results are stored in the classical register $\rm{c}0$ (In FIG. \ref{circuit}, "$\_$1" represents a classical register of 1 bit in length). If $\rm{c}0$ equals 1, the compensating operation $X$ and $R_X(\pi/2)H$ operation will be executed on $\rm{q}[1]$ circuit. If $\rm{c}0$ equals 0, $R_X(\pi/6)H$ operation will be executed on $\rm{q}[1]$ circuit. Next, we perform the $CZ$ entanglement operation between the $\rm{q}[1]$ and $\rm{q}[2]$ circuits, and perform the $H$ gate and the $\{|0\rangle,|1\rangle\}$ basis measurement operation on the $\rm{q}[1]$ circuit, which is equivalent to the measurement operation on the second particle in the $\{|+\rangle,|-\rangle\}$ basis. The measurement results are stored in the classical register $\rm{c}1$. If $\rm{c}1$ equals 1, the $X$ compensating operation and $R_X(5\pi/3)H$ operation will be performed on the $\rm{q}[2]$ circuit. If $\rm{c}1$ equals 0, the $R_X(-\pi/3)H$ operation will be performed on the $\rm{q}[2]$ circuit. At this point, we can obtain the quantum state $|\Psi'\rangle$ on the circuit $\rm{q}[2]$, as shown in Eq.(\ref{last-quantum-state}). Finally, we perform the $R_X(\pi/3)$ operation on $\rm{q}[2]$ to reconstruct the privacy quantum state $|\psi\rangle_A=\frac{1}{2}|0\rangle+\frac{\sqrt{3}}{2}|1\rangle$.

\begin{figure*}[!t]
	\centering
	\includegraphics[width=7 in]{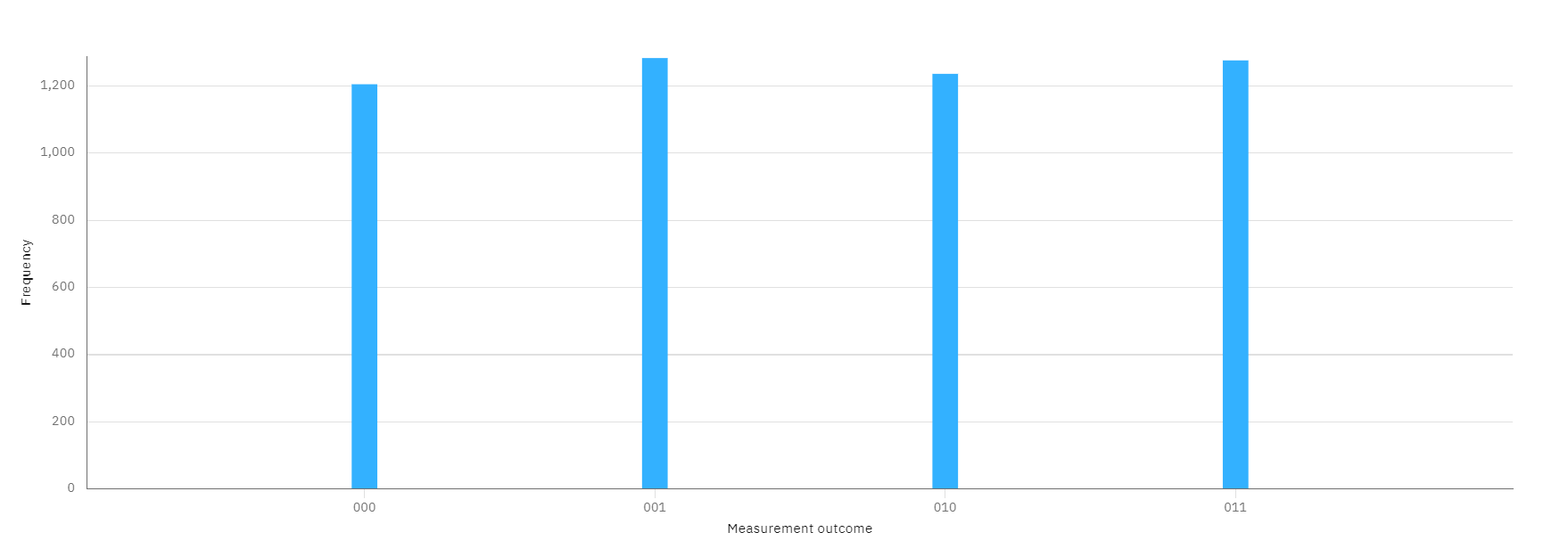}
	\caption{The results of the experiment.}
	\label{result}
\end{figure*}

Thirdly, to convert the reconstructed quantum state into a more convenient measurement result, we use $R_Y(-\pi/2)$, $R_Z(-\pi/3)$ and $H$ operations to transform it into a state that is more easily measured in the $\{|0\rangle,|1\rangle\}$ basis. If the reconstructed quantum state is indeed the original quantum state $|\psi\rangle$, then the state after transforming should be the $|0\rangle$ state. Regardless of how many times we measure it in the $\{|0\rangle, |1\rangle\}$ basis, each of the measurement result can only be 0, which is stored in the classical register $\rm{c}2$. Conversely, if it is not the original quantum state, we will obtain some measurement results of 1. The experimental measurement results are shown in FIG. \ref{result}. Each digit on the horizontal axis, such as 000, 001, 010 and 011, represents the results of the classical registers $\rm{c}2$, $\rm{c}1$ and $\rm{c}0$ respectively, one digit per register. The vertical axis represents the frequency of a certain measurement result. According to the results in FIG. \ref{result}, after a total of 5000 repeated experiments, we observed that the $\rm{c}2$ register was consistently 0, which was consistent with our expectations, namely that the recovered quantum state was equal to the original quantum state $|\psi\rangle$. Therefore, the experiment is successful.

\section{Application:Quantum Multiple Secrets Sharing}
Multiple secrets sharing (MSS) is a cryptographic technique that enables the sharing of multiple secrets among a group of participants, as opposed to just one single secret. Quantum multiple secrets sharing (QMSS) is an extension of MSS in the field of quantum information, allowing multiple quantum secrets to be shared among participants. In this section, we will show that the QMSS scheme can be implemented based on the presented protocol. 

\begin{figure*}[!t]
	\centering
	\includegraphics[width=7 in]{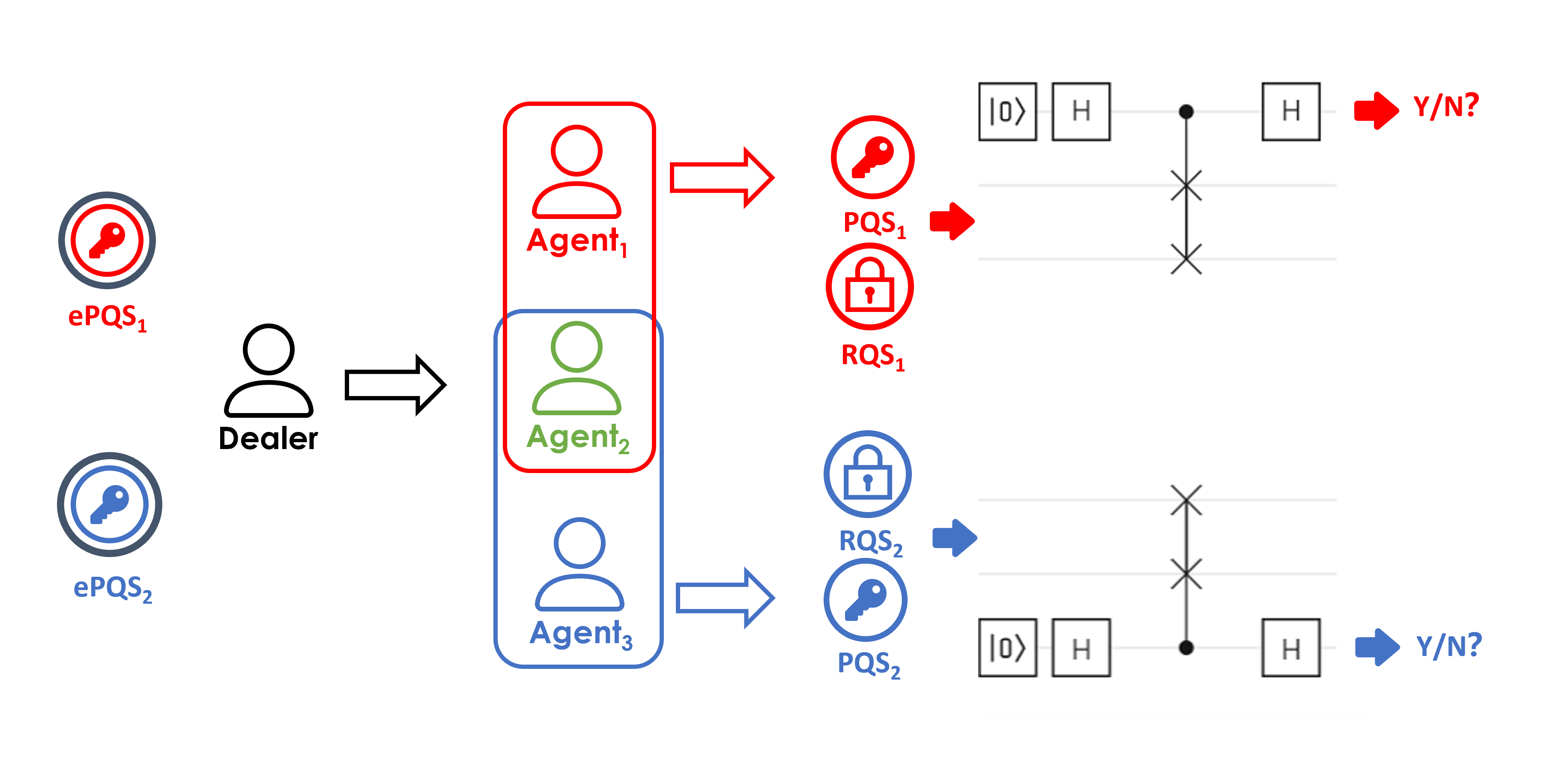}
	\caption{In our quantum multiple secrets sharing scheme, the dealer send multiple encrypted privacy quantum states(ePQS), such as ePQS$_1$ and ePQS$_2$, to the agents. The ePQS particles are the ones enclosed by the black circle which means the encryption. Then the agents collaborate with other agents to recover the privacy quantum states(PQS), such as PQS$_1$ and PQS$_2$. Finally, through the quantum circuit of fingerprint authentication, the combiners and the dealers confirm whether the PQS is identical to the raw quantum states (abbreviated as RQS), such as RQS$_1$ and RQS$_2$. }
	\label{QMSS}
\end{figure*}

In our QMSS scheme, the dealer first splits a secret using the classical $(t,n)$ threshold scheme and securely sends $n$ shares to agents (shareholders) through QKD. Then, the dealer encrypt $w$ different private quantum states (short for PQS) $|\psi_1\rangle,|\psi_2\rangle,\cdots,|\psi_w\rangle$  to encrypted private quantum states (short for ePQS) $|\Psi_1\rangle,|\Psi_2\rangle,\cdots,|\Psi_w\rangle$ and transmits them to different agents, where $w\geq2$. After receiving particle, each agent gathers a certain number of other agents, such as $t-1$, to reconstruct the private quantum states $|\psi_1\rangle,|\psi_2\rangle,\cdots,|\psi_w\rangle$, respectively. The above reconstruction process is similar to the reconstruction process of the proposed protocol. Finally, if the agent needs to verify the correctness of the private quantum state, he can use quantum fingerprinting \cite{Buhrman2001} for authentication with the dealer. It is worth nothing that one set of shares in our QMSS scheme is able to reconstruct multiple quantum private states. The diagram of the multiple secrets sharing scheme is shown in FIG. \ref{QMSS}.

Our protocol is simpler and more practical than the existing QMSS protocol, such as Ref. \cite{Samadder2022}. By comparison, we can find that the experimental implementation of \cite{Samadder2022} is relatively challenging, as it requires the use of trap codes and discrete quantum walks, which are more complicated technologies than cluster states. Although the authors provide mathematical descriptions and security analyses of the algorithm and protocol, there is a lack of actual experimental data to prove its feasibility and effectiveness. Furthermore, as Ref.\cite{Samadder2022} necessitates high-precision quantum operations and control, perfect hardware support is required, which poses a practical challenge. In comparison, our protocol based on cluster states has more feasibility and already been experimentally demonstrated. 

\section{Conclusion}
In this paper, we have shown how to address the security vulnerability in the  reconstruction process of quantum secret sharing protocols, which has never been effectively resolved in the previous papers. The idea of our paper is not to directly transmit shares, but rather to hide the shares within the particles. These particles are assembled in an appropriate manner. Ultimately, the privacy quantum state are recovered from these particles. Specifically, the dealer's particle comprises the information of privacy quantum state and her secret. Each shareholder's particle includes partial information of his share, and is transmitted in the public quantum channel. The other partial information is transmitted in the public classical channel. At the end of the protocol, it is worthwhile to note that only the privacy quantum state can be revealed, and all the other information, such as shares, remain perfectly secret. 

Our protocol is capable to be secure against various common internal and external attacks. Furthermore, we have successfully implemented the simulation experiment of our protocol on the quantum platform IBM Q, which proves the feasibility of the protocol. In addition, due to the protection of all shares during the reconstruction process, we have found that the protocol enables the reuse of shares, which is more efficiency compared to classical secret reconstruction protocols. Furthermore, this paper enables the realization of quantum multiple secrets sharing, which is simpler and more practical than the existing quantum multiple secrets sharing scheme.

\section{Acknowledgments}
This work was supported by National Natural Science
Foundation of China (Grants No. 62171131, 61976053,
and 61772134), Fujian Province Natural Science Foundation
(Grant No. 2022J01186), and Program for New Century Excellent Talents in Fujian Province University

\newpage 

\end{document}